\documentclass[american,english]{revtex4}
\usepackage[T1]{fontenc}
\usepackage[latin1]{inputenc}
\usepackage{babel}
\usepackage{textcomp}
\usepackage{amsthm}
\usepackage{amsmath}
\usepackage{graphicx}
\usepackage[unicode=true,pdfusetitle,
 bookmarks=true,bookmarksnumbered=false,bookmarksopen=false,
 breaklinks=false,pdfborder={0 0 1},backref=false,colorlinks=false]
 {hyperref}
\usepackage{breakurl}

\makeatletter
\@ifundefined{textcolor}{}
{%
 \definecolor{BLACK}{gray}{0}
 \definecolor{WHITE}{gray}{1}
 \definecolor{RED}{rgb}{1,0,0}
 \definecolor{GREEN}{rgb}{0,1,0}
 \definecolor{BLUE}{rgb}{0,0,1}
 \definecolor{CYAN}{cmyk}{1,0,0,0}
 \definecolor{MAGENTA}{cmyk}{0,1,0,0}
 \definecolor{YELLOW}{cmyk}{0,0,1,0}
 }
\numberwithin{equation}{section}
\numberwithin{figure}{section}

\usepackage{graphicx}
\usepackage{amssymb}

\usepackage{babel}

\makeatother

\begin{document}

\title{A Formalism for Scattering of Complex Composite Structures. 2 Distributed
Reference Points}

\author{Carsten Svaneborg$^{1,2*}$ and Jan Skov Pedersen$^{2}$}

\affiliation{$^{1}$Center for Fundamental Living Technology, Department of Physics,
Chemistry and Pharmacy, University of Southern Denmark, Campusvej
55, DK-5320 Odense, Denmark}

\affiliation{$^{2}$Department of Chemistry and Interdisciplinary Nanoscience
Center (iNANO), University of Aarhus, Langelandsgade 140, DK-8000
Århus, Denmark.}
\begin{abstract}
Recently we developed a formalism for the scattering from linear and
acyclic branched structures build of mutually non-interacting sub-units.{[}C.
Svaneborg and J. S. Pedersen, J. Chem. Phys. 136, 104105 (2012){]}
We assumed each sub-unit has reference points associated with it.
These are well defined positions where sub-units can be linked together.
In the present paper, we generalize the formalism to the case where
each reference point can represent a distribution of potential link
positions. We also present a generalized diagrammatic representation
of the formalism. Scattering expressions required to model rods, polymers,
loops, flat circular disks, rigid spheres and cylinders are derived.
and we use them to illustrate the formalism by deriving the generic
scattering expression for micelles and bottle brush structures and
show how the scattering is affected by different choices of potential
link positions.
\end{abstract}
\maketitle

\section{Introduction}

Light scattering, small-angle neutron or X-ray scattering (LS, SANS
and SAXS, respectively) techniques are ideal for obtaining detailed
information about self-assembled molecular and colloidal structures.\cite{GuinierFournet,HigginsBenoit,LindnerZemb}
However, these techniques provide reciprocal space intensity spectra.
Typically such spectra can not be interpreted directly, instead extensive
modeling is required to infer structural information. Such an analysis
necessitates the availability of a large tool-box of model expressions
characterizing the scattering spectra expected from many different
well defined structures. Fitting such expressions to experimental
scattering spectra allows the experimentalist to infer and accurately
quantify which structures are most likely to be present in a given
sample.

While scattering theory and statistical mechanics provides a general
framework for how to derive such models, there are no general analytical
methods for deriving the expected scattering spectra for complex self-assembled
structures. Often unchecked approximations need to be introduced to
obtain analytical results. Alternatively computer simulations can
be employed to make virtual scattering experiments from ensembles
of well defined structures, which can then be compared to the experimental
scattering spectra. Our aim here is to present a formalism for deriving
the scattering expressions characterizing a large class of structures. 

We assume the structures are build out of well defined components
which we call sub-units. We make no assumptions as to the internal
structure a sub-units, nor on the number of sub-unit types that can
be present in a given structure. Sub-units have well defined reference
points by which they can be joined to other sub-units. The formalism
in its present form requires that all such joints are completely flexible.
Finally the formalism requires that the structures formed by joining
sub-units does not contain loops. For structures that meet these requirements,
the formalism allows exact scattering expressions characterizing the
corresponding scattering spectrum to be derived with great ease. The
central idea is to express the scattering of the whole structure in
terms of scattering expressions characterizing the sub-units instead
of the scattering from the individual scattering sites comprising
the structure. This idea was previously used by Benoit and Hadziioannou\cite{BenoitHadziioannouMacromolecules88}
to calculate the scattering from various block-copolymer structures,
and by Read\cite{ReadMacromolecules98} who applied it to calculate
the scattering from H-polymers and stochastic branched polymer structures.
Teixeira et al. used this idea to calculate the scattering from structures
composed of polymer/rod polycondensates\cite{TeixeiraMM2000,TeixeiraJCP2007}.

In a previous paper\cite{cs_jpc_submitted1}, we derived and presented
a versatile formalism for predicting the scattering from linear and
branched structures composed of arbitrary functional sub-units. We
argued that the formalism is complete in the following sense: Three
functions describe the scattering from a sub-unit, and we derive three
analogous scattering expressions that describe the scattering from
a structure. Hence, we can 1) build bottom-up hierarchical structures
by building structures by joining sub-units to well defined sites
in a structure and 2) build top-down hierarchical structures by substituting
sub-units by more complex sub-structures composed of sub-units. Furthermore
the formalism is generic, in the sense that scattering contributions
from structural connectivity and the internal sub-unit structures
are decoupled. This allows generic structural scattering expressions
to be derived and composed to describe the scattering from complex
hierarchical structures independently of which sub-units we build
the structure. We also developed a diagrammatic interpretation of
the formalism that allows us to map structural transformations onto
algebraic transformations of the corresponding scattering expressions.
We illustrated this with the transformation rules producing the scattering
expression for an $n$'th generation dendrimer, by successive replacement
of the outermost leafs by star shaped sub-structures. In this previous
paper\cite{cs_jpc_submitted1}, we focused on structural complexity
and illustrated the formalism by deriving the structural scattering
expressions of linear chain structures, stars, pom-poms, bottle-brushes
(i.e. chains of stars) and dendritic structures (i.e. stars build
of stars). 

In the present paper our focus is to present the expressions characterizing
a large variety of possible sub-units such as rigid rods, flexible
and semi-flexible polymers, loops of flexible polymers, and excluded
volume polymers. We also present general expressions for geometric
sub-units together with the expressions for special cases such as
disks, spherical shells, solid spheres and cylinders. While the form
factors for most of the sub-units are well known, the form factor
amplitudes and phase factors depend crucially on how we choose to
link the sub-units together. Different choices of linking (for instance
center-to-center or surface-to-surface) cause additional geometric
factors to appear in the form factor amplitudes and phase factors.
To illustrate the formalism, we use it to predict the scattering for
chains of identical sub-units or alternating sub-units. Furthermore,
we address the situation where a single sub-unit can have a distribution
of reference points, and hence form factor amplitudes and phase factors
need to be averaged over linking probability distributions. Together,
paper I and the present paper allow the scattering expressions for
complex heterogeneous structures of a variety of sub-units to be derived
with great ease.

The paper is structured as follows: In Sect. \ref{sec:Theory} we
present the formalism and generalize it for structures where reference
points can be distributed. We illustrate the formalism by deriving
the generic scattering for a block copolymer micelle. In Sect. \ref{sec:Linear}
we present the general scattering expressions characterizing an arbitrary
linear sub-unit with internal conformational degrees of freedom, and
in Sect. \ref{sec:Linear-examples} we give examples of the scattering
from chains and bottle-brush structures. In Sect. \ref{sec:Geometric}
we present the general scattering expressions characterizing an arbitrary
geometrical sub-unit without internal conformations, and in Sect.
\ref{sec:geometric-examples} we give examples of the scattering from
block copolymer micelles with different core sub-units and tethering
geometries as well as the scattering from end-linked cylinders. We
present our conclusions in Sect. \ref{sec:Conclusions}. In two Appendices,
we derive the scattering terms for polymers, rods, and closed polymer
loops, and for spheres, disks and cylinders taking different tethering
geometries into account.

\section{Theory\label{sec:Theory}}

The present theory pertains to the small-angle scattering for structures
build out of sub-units and how to efficiently derive the scattering
spectra characterizing such structures. The formalism is identical
for light, X-ray or neutron scattering experiments within the Rayleigh-Debye-Ganz
approximation. We define an excess scattering length for each sub-unit.
This parameter captures the experimental details of the interactions
between the incident radiation and the scatterers inside the sub-unit,
and also the scattering properties of the solvent in which we assume
the structures are dissolved.

Each sub-unit comprises a specific number of scattering sites. We
equip each sub-unit with an arbitrary number of reference points,
these are positions on the sub-unit where we can join two or more
sub-units together. Later we will generalize each reference point
to represent a distribution of such positions. If the sub-unit is
a polymer molecule, then a natural choice could be to have the two
ends as reference points, if we are interested in deriving the scattering
from end-linked polymer structures. Assume that the $I$'th sub-unit
is composed of point-like scatterers, where the $j$'th scatterer
in the sub-unit is located at a position ${\bf r}_{Ij}$ and has excess
scattering length $b_{Ij}$. Let ${\bf R}_{I\alpha}$ denote the position
of the $\alpha$'th reference point associated with the $I$'th sub-unit.
Once two or more sub-units are connected at the same reference point,
we refer to it as a vertex in the resulting structure, e.g. if sub-units
$I$ and $J$ are joined at reference point $\alpha$ then ${\bf R}_{I\alpha}={\bf R}_{J\alpha}$
denotes the same position in space and a vertex in the structure.
Here and in the following capital letters refers to sub-units, lower
case letters refers to scatterers inside a sub-unit, and Greek letters
refers to vertices and reference points.

Scattering experiments measures the distribution of pair-distances
between scatterers in a structure. For a given structure $S$ we can
define three types of pair-distance distributions. The form factor
$F_{S}(q)$ is the excess scattering length weighted Fourier transformed
and conformationally averaged pair-distance distribution between all
scatterers in the structure; this is what is measured in a scattering
experiment. We can also define two auxiliary pair-distance distributions.
The form factor amplitude $A_{S\alpha}(q)$ which is the scattering
length weighted Fourier transform of the pair-distance distribution
between all scatterers in the structure and a specified vertex $\alpha$.
Finally, the phase factor $\Psi_{S\alpha\omega}$ is the Fourier transform
of the pair-distance distribution between two vertices $\alpha$ and
$\omega$ in the structure.

We can define the form factor, form factor amplitudes and phase factors
of a structure $S$ in terms of the scattering sites and reference
points as

\begin{equation}
F_{S}(q)=\left(\beta_{S}\right)^{-2}\left\langle \sum_{j,k}b_{Sj}b_{Sk}e^{i{\bf q}\cdot({\bf r}_{Sj}-{\bf r}_{Sk})}\right\rangle _{S},\label{eq:FI}
\end{equation}

\begin{equation}
A_{S\alpha}(q)=\left(\beta_{S}\right)^{-1}\left\langle \sum_{j}b_{Sj}e^{i{\bf q}\cdot({\bf r}_{Sj}-{\bf R}_{S\alpha})}\right\rangle _{S},\label{eq:AI}
\end{equation}
and

\begin{equation}
\Psi_{S\alpha\omega}(q)=\left\langle e^{i{\bf q}\cdot({\bf R}_{S\alpha}-{\bf R}_{S\omega})}\right\rangle _{S}.\label{eq:PI}
\end{equation}

The $\langle\text{\ensuremath{\cdots}}\rangle_{S}$ averages are over
internal conformations and orientations. The total scattering length
of the whole structure is $\beta_{S}=\sum_{j}b_{Sj}$. Due to the
orientational average, all the functions only depend on the magnitude
of the momentum transfer $q$, which is given by the angle between
the incident and scattered beam and the wave length of the radiation.
We also have $\Psi_{S\alpha\omega}=\Psi_{S\omega\alpha}$ due to the
orientational average. Here and in the rest of the paper, the form
factor, form factor amplitudes, and phase factors are normalized to
unity in the limit $q\rightarrow0$.

The derivation of scattering expressions for complex structures can
be vastly simplified by describing the structure not in terms of fundamental
scattering sites, but instead in terms of logical structural sub-units
of the structure. Each sub-unit corresponds to a well defined group
of the scattering sites, and is characterized by its own form factor,
form factor amplitudes, and phase factors defined by eqs. \ref{eq:FI}-\ref{eq:PI}.
For instance, to derive the scattering expression for a block-copolymer
micelle, we can group all the scattering sites of the core into one
sub-unit, and let each polymer in the corona be described by a sub-unit.

The fundamental result of the present formalism, is to express the
form factor, form factor amplitudes and phase factors of a whole structure
in terms of the same functions characterizing the sub-units. An exact
and generic expression can only be derived in the case where the internal
conformations of all sub-units are uncorrelated, since in this case
the structural average factorizes into single-sub unit averages. This
allows generic scattering expressions to be derived for a large class
of complex structures. The assumption of uncorrelated sub-units corresponds
to assuming that sub-units are mutually non-interacting, that joints
are completely flexible, and that the structure does not contain loops.
Subject to these assumptions, we can succinctly express the form factor,
form factor amplitudes and phase factors of a structure $S$ as

\begin{equation}
F_{S}(q)=\beta_{S}^{-2}\left[\sum_{I}\beta_{I}^{2}F_{I}(q)+\sum_{\substack{I\neq J\\
\alpha\in I\:\mbox{near}\:\omega\in J
}
}\beta_{I}\beta_{J}A_{I\alpha}(q)A_{J\omega}(q)\prod_{\substack{(K,\tau,\eta)\\
\in\mbox{P}(\alpha,\omega)
}
}\Psi_{K\tau\eta}(q)\right],\label{eq:F}
\end{equation}

\begin{equation}
A_{S\alpha}(q)=\beta_{s}^{-1}\left[\sum_{\substack{I\\
\omega\in I\,\mbox{near}\,\alpha
}
}\beta_{I}A_{I\omega}(q)\prod_{\substack{(K,\tau,\eta)\\
\in\mbox{P}(\alpha,\omega)
}
}\Psi_{K\tau\eta}(q)\right],\label{eq:A}
\end{equation}
and

\begin{equation}
\Psi_{S\alpha\omega}(q)=\prod_{\substack{(K,\tau,\eta)\\
\in\mbox{P}(\alpha,\omega)
}
}\Psi_{K\tau\eta}(q).\label{eq:P}
\end{equation}

Here $F_{I}$ denotes the form factor of the $I$'th sub-unit, $A_{I\alpha}$
denotes the form factor amplitude of the $I$'th sub-unit relative
to the reference point $\alpha$, $\Psi_{I\tau\eta}$ denotes the
phase factor of the $I$'th sub-unit between reference points $\tau$
and $\eta$, and$\beta_{I}=\sum_{j}b_{Ij}$ the total excess scattering
length of the $I$'th sub-unit. These terms are defined as eqs. \ref{eq:FI}-\ref{eq:PI}
with $S$ replaced by $I$. In the form factor we have a double sum
over distinct sub-unit pairs, and in the form factor amplitude a single
sum over sub-units. In the form factor sum, the restriction $\alpha\in I$
near $\omega\in J$ means that we for have to identify the reference
point $\alpha$ on $I$ nearest to $J$ in terms of the structural
connectivity, and similarly the reference point $\omega$ on $J$
nearest to $I$. Having done this, we can identify the path $P(\alpha,\omega)$
of sub-units ($K$) and reference point pairs ($\tau$,$\eta$) that
has to be traversed to walk between the $I$'th and $J$'th sub-unit
to from the reference point $\alpha$ to the reference point $\omega$
on the structure. Since we are assuming acyclic branched structures
this path definition is always unique. Details of the derivation of
this expression is given in ref. \cite{cs_jpc_submitted1}. In the
example section below, we will show how to derive the scattering for
a few concrete structures using eqs. \ref{eq:F}-\ref{eq:P}. First
we will generalize the formalism to the case of random linking positions,
and present a diagrammatic illustration of the physical interpretation
of the generalized formalism.

Above we have assumed that sub-units are always linked at reference
points which correspond to specific sites on the sub-units. In the
following, we refer to this as regular reference points. For many
structures there is an element of randomness to where sub-units are
joined together. For an example of a bottle-brush structure, we can
for instance imagine a main rod with polymer sub-units linked at random
positions along the rod. While the structure has an element of randomness,
the connectivity remains well-defined. Note, that this situation differs
from random linking, where the structure will have a random connectivity.
Here and below we only address the first situation. Random linking
can be described by cascade theory\cite{GordonProcRoySoc1962,Burchard}
or Markov chain models\cite{TeixeiraMM2000,TeixeiraJCP2007}. 

The formalism can easily be generalized to the case where some or
all reference points refers to distributions of potential link positions
on a sub-unit. Above, we have assumed that a reference point $\alpha$
on a sub-unit refers to a unique fixed position ${\bf R}_{I\alpha}$
on the sub-unit, and that sub-unit pairs $I$ and $J$ are joined
at the $\alpha$ vertex when ${\bf R}_{I\alpha}={\bf R}_{J\alpha}$.
Below we consider the situation where a reference points can be picked
from a given distribution. In this case we refer to the reference
point as an distributed reference point. Assuming we are given a set
of potential positions for the $\alpha$th reference point on sub-unit
$I$ where the $m$th possible position is ${\bf R}_{I\alpha m}$
and is associated with a probability $Q_{I\alpha m}$ and similar
${\bf R}_{J\alpha n}$ and $Q_{J\alpha n}$ for the $n$th possible
position of the $\alpha$th reference point on sub-unit $J$. Then
the probability of joining two sub-units $I$ and $J$ at the $\alpha$
vertex at specific positions ${\bf R}_{I\alpha m}={\bf R}_{J\alpha n}$
is given by the product $Q_{I\alpha m}Q_{J\alpha n}$. This is tantamount
to assuming that the two linking positions on the two joined sub-units
are statistically uncorrelated. Note that one or both of these distributions
can still refer to a single potential position, hence regular reference
points remains a special case of the generalized formalism.

To derive eqs. \ref{eq:F}-\ref{eq:P}, we had to assume that the
internal conformations of all sub-units were uncorrelated. This allowed
structural averages to be factorized into single-sub unit averages.
When we have assigned a probability distribution to some or all reference
points we have to calculate $\langle A_{S\alpha}(q)\rangle_{{\bf Q}}$,
$\langle\Psi_{S\alpha\omega}(q)\rangle_{{\bf Q}}$ where $\langle\cdots\rangle_{{\bf Q}}$
denotes the additional averages over link position distributions.
Since we have assumed that linking positions on different sub-units
are statistically independent, the structural and linking position
averages again factorize into single sub-unit averages over internal
conformations as well as linking degrees of freedom. Sub-unit form
factors are independent of reference points, and hence unaffected
by this average. For a sub-unit with a finite set of potential random
linking positions, we can define the reference point averaged form
factor amplitudes and phase factors as

\begin{equation}
A_{I\langle\alpha\rangle}(q)=\left\langle A_{I\alpha}(q;{\bf R}_{I\alpha})\right\rangle _{Q_{I\alpha}}\equiv\sum_{n}Q_{I\alpha n}A_{I\alpha}(q,{\bf R}_{I\alpha n}),\label{eq:aaverage}
\end{equation}

\begin{equation}
\Psi_{I\langle\alpha\rangle\langle\omega\rangle}(q)=\left\langle \Psi_{I\alpha\omega}(q;{\bf R}_{I\alpha},{\bf R}_{I\omega})\right\rangle _{Q_{I\alpha}Q_{I\omega}}\equiv\sum_{n,m}Q_{I\alpha n}Q_{I\omega m}\Psi_{I\alpha}(q;{\bf R}_{I\alpha n},{\bf R}_{I\omega m}).\label{eq:paverage}
\end{equation}

\begin{figure}
\includegraphics[scale=0.5]{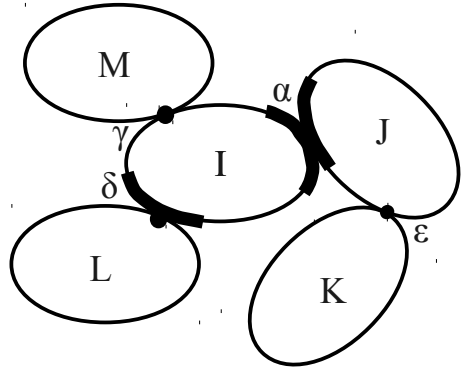}

\caption{\label{fig:Example-structure}Illustration of an example structure
where five sub-units have been joined at four vertices $\alpha$,
$\gamma$, $\delta$, $\epsilon$. Each sub-unit is illustrated by
an ellipse where we associate the interior with the internal conformation
of the scattering sites and the circumference with the reference points
on the sub-unit. Sub-units are joined to each other by reference points,
and they are illustrated as a single dot (e.g. $\epsilon$, $\delta$,
$\gamma$) for a regular reference point or a thick line (e.g. $\langle\alpha\rangle$,
$\langle\delta\rangle$) for distributed reference points. Note that
we use the same Greek letter to label vertices in the structure, and
regular or distributed reference points on different sub-units at
the same vertex.}
\end{figure}

Here we write explicitly the reference points, that are to be averaged
over in the form factor amplitudes and phase factors. Here and below
we use $\langle\alpha\rangle$ to denote the case where the $\alpha$th
reference point label on the $I$th sub-unit has been averaged. We
continue to denote by $\alpha$ a regular reference point, where no
average is to be performed. The formalism (eqs. \ref{eq:F}-\ref{eq:P})
remains valid when some or all form factor amplitudes and phase factors
include distributed reference points.

Fig. \ref{fig:Example-structure} schematically illustrates an example
structure for which the formalism can provide the corresponding scattering
expression. It shows how sub-units can be joined either at regular
reference points or at distributed reference points. Note how a reference
point is associated with a sub-unit, hence $\delta$ refers to a regular
reference point on sub-unit $L$, that is joined to any of the $\delta$
reference points on the $I$th sub-unit denoted by the $\langle\delta\rangle$
average reference point. In general, a vertex have an arbitrary functionality,
and a sub-unit can have an arbitrary number of reference points at
which it can join with other sub-units. Hence the structures described
by formalism is not limited to two-functional graph-like structures,
but to any hyper-graph structures that does not contain loops.

In the special case, where we can make a one-to-one identification
between reference points and scattering sites such that $n=i$, ${\bf R}_{I\alpha n}={\bf r}_{Ii}$,
and $Q_{I\alpha n}=b_{Ii}/\beta_{I}$. Then \ref{eq:aaverage}, \ref{eq:AI},
\ref{eq:paverage}, and \ref{eq:PI} are identical to the form factor
eq. \ref{eq:FI}. Hence we conclude that $A_{I\langle\alpha\rangle}(q)=\Psi_{I\langle\alpha\rangle\langle\omega\rangle}(q)=F_{I}(q)$
in this case. For a polymer chain, for instance, this means that the
form factor amplitude relative to a random position on the polymer,
and the phase factor relative to two random positions on the polymer
are identical to the polymer form factor. This is not a surprise since
the site-to-site, site-to-reference point and reference-to-reference
point pair-distance distributions all are identical in this case.
When deriving form factors for a given structure, we often assume
that all sites in a sub-unit has equal excess scattering length, hence
all the known expressions for form factors of structures can be uses
as reference point averaged form factor amplitudes and phase factors
to describe structures with distributed link positions.

\begin{figure}
\includegraphics[scale=0.5]{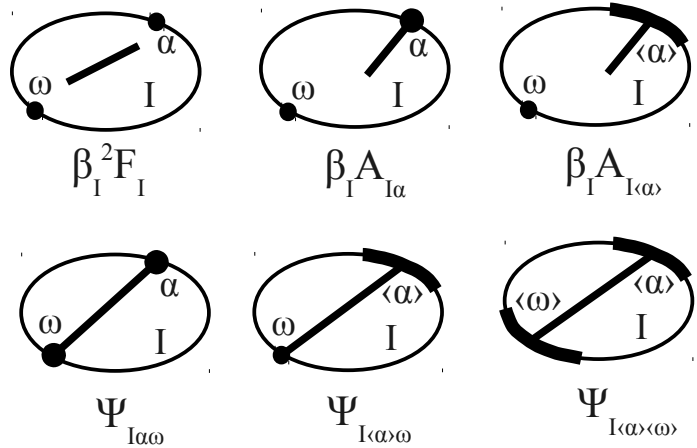}

\caption{\label{fig:Definition-of-diagrams}Definition of diagrams representing
the sub-unit form factor, form factor amplitude, and phase factor
terms expressed for the different possibilities of reference point
averages for the $I$th sub-unit.}
\end{figure}

Fig. \ref{fig:Definition-of-diagrams} introduces a diagrammatic interpretation
of the form factor, form factor amplitude and phase factor of a sub-unit.
As in fig. \ref{fig:Example-structure} we associate each sub-unit
with an ellipse, where reference points are associated with the circumference
while the scattering sites are associated with the interior. The form
factor is the Fourier transform of the pair-distance distribution
between all scattering sites in a sub-unit, and we illustrate this
by a straight line inside the sub-unit ellipse. Form factor amplitudes
and phase factors depend on reference points. Regular reference points
are shown as dots, while distributed reference points are illustrated
as a thick line on the circumference. The form factor amplitude is
Fourier transform of the pair-distance distribution between a specified
reference point and all scattering sites inside the sub-unit, and
this is illustrated as a line from dot to the inside of the sub-unit.
In the case, of a reference point average, we illustrate the reference
point not as a dot but by a thick line illustrating all the possible
reference points, and the form factor amplitude as a line from anywhere
along the thick line to the inside of the sub-unit. The phase factor
is the Fourier transform of the pair-distance distribution between
two specified reference points, and this is illustrated as a line
traversing the sub-unit connecting two reference points or reference
point averages. 

Using eq. \ref{eq:FI}, we can calculate the scattering form factor
for a given structure composed of sub-units joined by regular or distributed
reference points. The first term is just a sum over the form factors
of all the sub-units weighted by their excess scattering lengths.
The second term is more complicated, and it describes the scattering
interference contributed by different sub-unit pairs. For each distinct
pair of sub-units $I$ and $J$ in the double sum, we identify which
vertex $\alpha$ at sub-unit $I$ that is nearest sub-unit $J$ and
which vertex $\omega$ at sub-unit $J$ that is nearest to sub-unit
$I$. Here {}``near'' means in terms of the shortest path originating
at a reference point $\alpha$ (or $\langle\alpha\rangle$) on $I$
and terminating on a reference point $\omega$ (or $\langle\omega\rangle$)
on $J$. We denote the path connecting $\alpha$ and $\omega$ through
the structure $P(\alpha,\omega)$. For the product, we have to identify
each sub-unit $K$ on this path and also identify the reference points
$\tau$ and $\eta$ (or $\langle\tau\rangle$, $\langle\eta\rangle$)
across which the sub-unit is traversed. For some structures the path
can traverse a sub-unit by the same reference point. In the case of
a well defined reference point $\Psi_{K\alpha\alpha}(q)\equiv1$ and
we can neglect the contribution, however in the case of an distributed
reference point the corresponding term $\Psi_{K\langle\alpha\rangle\langle\alpha\rangle}(q)$
will contribute to the product. The path construction is always unique
and well defined for structures that does not contain loops.

The form factor expression (eq. \ref{eq:F}) has a quite simple physical
interpretation, despite the complex notation required to describe
branched reference point distributed structures. The structural form
factor is the pair-correlation function between all scattering sites
in the structure. It can be obtained by propagating position information
between all scattering sites in the structure. When both sites belong
to the same sub-unit it is given by the sub-unit form factors and
is described by the first term in eq. \ref{eq:F}. The distance information
between sites on different sub-units is obtained by propagating position
information along paths through the structure. To propagate site-to-site
position information between sites in sub-unit $I$ and sites in sub-unit
$J$, we first have to propagate the position information between
the sites in sub-unit $I$ to the vertex $\alpha$ at $I$. This is
done by the form factor amplitude $\beta_{I}A_{I\alpha}$ or $\beta_{I}A_{I\langle\alpha\rangle}$.
The position information is then propagated step-by-step along the
path of intervening sub-units towards the vertex $\omega$ at sub-unit
$J$. Each time a sub-unit is traversed it contributes a phase factor
$\Psi_{K\tau\eta}$, $\Psi_{K\langle\tau\rangle\eta}$,$\Psi_{K\tau\text{\ensuremath{\langle}}\eta\rangle}$,
or $\Psi_{K\langle\tau\rangle\langle\eta\rangle}$ to account for
the conformationally averaged distance between the two (potentially
distributed) reference points. Finally the position information is
propagated between the vertex $\omega$ and all the sites inside the
$J$ sub-unit. This is done by the final form factor amplitude $\beta_{J}A_{J\omega}$
or $\beta_{J}A_{J\langle\omega\rangle}$. Only the amplitudes has
an excess scattering length prefactor, since they represent the amplitudes
of scattered waves from all the scatterers inside the sub-units relative
to the $\alpha$ and $\omega$ vertices while the product of phase
factors represent excess phase contributed by the path between the
vertices. The product of all these propagators describe the scattering
length weighted site-to-site scattering interference contribution
from the $I$'th and $J$'th sub-units. By summing over all such pair
contributions all the possible site-to-site pair-distances in the
structure are accounted for.

\begin{figure}
\includegraphics[scale=0.5]{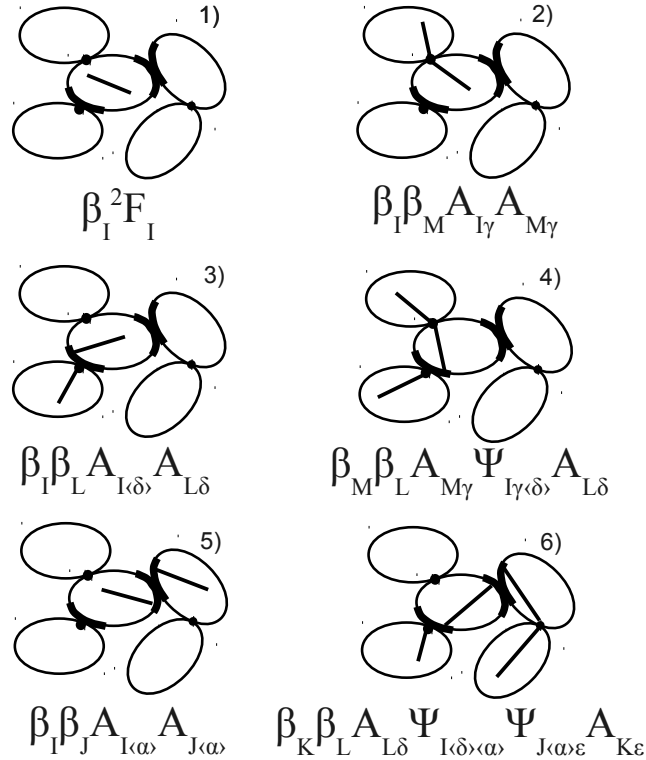}

\caption{\label{fig:Example-terms}Diagrammatic representation of some of the
terms contributing to the form factor of the structure shown in fig.
\ref{fig:Example-structure} using the definition of diagrams shown
in fig. \ref{fig:Definition-of-diagrams}.}
\end{figure}
Fig. \ref{fig:Example-terms} diagrammatically illustrates some of
the terms that contributes to the form factor (eq. \ref{eq:F}) using
the diagrammatic definitions in fig. \ref{fig:Definition-of-diagrams}.
We can generate all the diagrams by picking a pair of sites inside
the same or two different sub-units and drawing a line between them
using reference points to step between sub-units and to traverse across
sub-units (diagram 1). A line between two sites within the same sub-unit
contributes the form factor of that sub-unit. Sub-units that are joined
directly to each other will contribute the product of two form factor
amplitudes and excess scattering lengths of the two sub-units (diagram
2, 3, 5). For sub-units that are not directly joined to each other,
the form factor amplitude product is further multiplied by the phase
factors of all the sub-units on the intervening path (diagram 4, 6).
For all the reference point labels of the form factors and phase factors,
we either have a regular reference point, or a distributed reference
point which depend on the details of the given structure. In general,
for a structure of $N$ sub-units, there will be $N$ form factor
contributions and $N(N-1)/2$ different scattering interference contributions
that has to be determined. The longest possible path is $N-2$ sub-units
which occurs in the case of a linear chain of sub-units.

Similar diagrammatic interpretations apply to the structural form
factor amplitude and phase factors (eqs. \ref{eq:A} and \ref{eq:P}).
For the form factor amplitude we have to propagate position information
between a specified vertex and all sites in the structure. Diagrammatically
this can be done by picking a site in any sub-unit and drawing a line
between the site and the specified vertex using reference points to
step between sub-units and to traverse across sub-units. Summing over
all the $N$ diagrams will produce the form factor amplitude of the
structure. For the phase factor we have to propagate position information
between two specified vertices. Diagrammatically this is done by drawing
a line between the two vertices using reference points to step between
sub-units and to traverse across sub-units. The resulting structural
phase factor is the product of the phase-factors of all the intervening
sub-units.

With the diagrammatic interpretation of the formalism, it becomes
quite easy to draw a structure, and write down the corresponding scattering
expressions. The price of this simplicity is that we had to assume
that sub-units are mutually non-interacting, that the joints between
sub-units are completely flexible, and that the structures does not
contain loops. However, no assumptions were made about the internal
structure of the sub-units. The formalism is complete in the sense
that the three structural scattering expressions allows a whole structure
to be used as a sub-unit to build more complex structures. This we
utilized in Paper 1 but will not use here.\cite{cs_jpc_submitted1}
The formalism is also generic in the sense that scattering contributions
due to structural connectivity and the internal structure of the sub-units
have been completely decoupled. This allows us to write down generic
scattering expressions for structures without knowing what sub-units
they are made of. This information can be specified at a later point
when concrete expressions for the sub-unit form factor, form factor
amplitudes, and phase factors are inserted. Below we give some generic
examples, and then dedicate the rest of the paper to derive and present
scattering expressions characterizing specific sub-unit structures.

\subsection{Example structures\label{sub:Micelle-example}}

\begin{figure}
\includegraphics[scale=0.5]{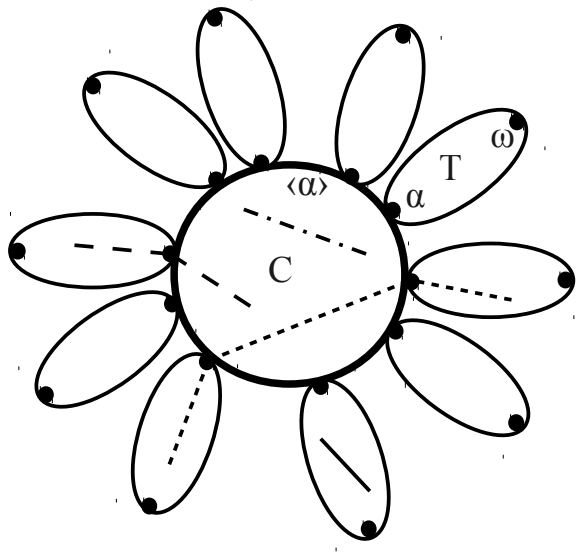}\caption{Diagrammatic representation of the form factor of a block copolymer
micelle with a core sub-unit {}``C'' and a number of identical sub-units
{}``T'' tethered at random points on the surface. All the scattering
contributions are shown using the diagrammatics in fig. \ref{fig:Definition-of-diagrams}.
They correspond to $\beta_{T}^{2}F_{T}$ (solid line), $\beta_{T}^{2}A_{T\alpha}^{2}\Psi_{C\langle\alpha\rangle\langle\alpha\rangle}$
(short dashed line), $\beta_{T}\beta_{C}A_{T\alpha}A_{C\langle\alpha\rangle}$
(long dashed line), and $\beta_{C}^{2}F_{C}$ terms (dot dashed line)
in eq. \ref{eq:f_micelle}. \label{fig:micelle}}
\end{figure}
\begin{figure}
\includegraphics[scale=0.5]{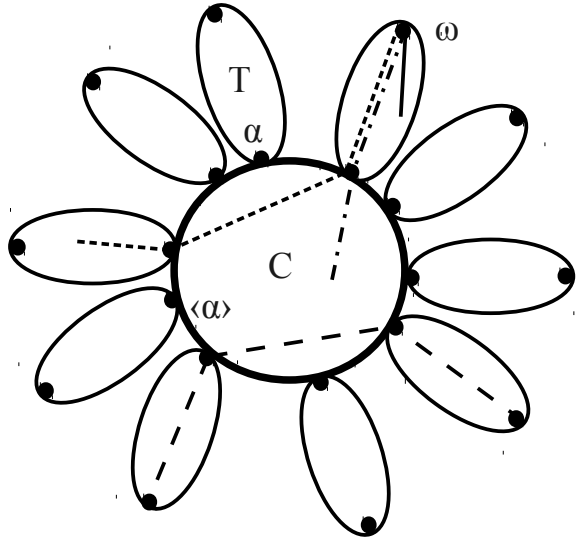}\caption{Diagrammatic representation of the form factor amplitude and phase
factor of a block copolymer micelle. The scattering contributions
to $A_{mic,\omega}$ are shown by lines starting at an $\omega$ reference
point on a tethered sub-unit and using the diagrammatics in fig. \ref{fig:Definition-of-diagrams}.
They correspond to the $\beta_{T}A_{T\omega}$ (solid line), $\beta_{C}A_{C\langle\alpha\rangle}\Psi_{T\omega\alpha}$
(dash dotted line), and $\beta_{T}A_{T\alpha}\Psi{}_{C\langle\alpha\rangle\langle\alpha\rangle}\Psi_{T\alpha\omega}$
terms (short dashed line) in eq. \ref{eq:a_micelle}. The phase factor
(eq. \ref{eq:p_micelle}) between the tips of two tethered sub-units
is given by $\Psi_{T\omega\alpha}^{2}\Psi_{C\langle\alpha\rangle\langle\alpha\rangle}$
(long dashed line) . \label{fig:micelleAP}}
\end{figure}
We can regard a micelle as $N$ identical two-functional sub-units
tethered by one end to a random site on the surface of geometric structure
representing the core. Such a structure is shown in fig. \ref{fig:micelle}.
Note that the core surface still is referred to by a single reference
point label $\alpha$. Similarly, we can regard a bottle-brush polymer
as $N$ identical two-functional sub-units tethered by one end to
a random point on a main structure such as a polymer chain. The connectivity
of the two structures is identical, and hence they are characterized
by the same generic scattering expression:

\[
F_{mic}(q)=\left(N\beta_{T}+\beta_{C}\right)^{-2}\left(\beta_{C}^{2}F_{C}+N\beta_{T}^{2}F_{T}\right.
\]

\begin{equation}
\left.+2N\beta_{C}\beta_{T}A_{C\langle\alpha\rangle}A_{T\alpha}+N(N-1)\beta_{T}^{2}A_{T\alpha}^{2}\psi_{C\langle\alpha\rangle\langle\alpha\rangle}\right)\label{eq:f_micelle}
\end{equation}

Here the tethered sub-units are denoted by $T$ and the end attached
to the core surface is denoted {}``$\alpha$'' while {}``$\omega$''
denotes the free end. The core sub-unit is denoted by $C$ and the
average over random surface points is denoted by $\langle\alpha\rangle$.
The form factor consists of terms representing all the possible pair
distributions between sub-units in the structure. These are shown
in fig. \ref{fig:micelle}. Each term has a prefactor which for form
factor terms is the number of corresponding sub-units in the structure.
The form factor amplitude product terms represent pair distributions
between different sub-units and they are counted twice. Hence there
is both an $A_{C\langle\alpha\rangle}A_{T\alpha}$ contribution and
an identical $A_{T\alpha}A_{C\langle\alpha\rangle}$ contribution
for each of the $N$ tethered sub-units. For the pair distribution
between two tethered sub-units, we note that we have $N$ tethered
sub-units to pick the starting point from, and $N-1$ tethered sub-units
to pick end ending point. This also counts each pair twice. The prefactor
of the form factor ensures it is normalized to unity in the limit
of small $q$ values.

We have chosen to express the form factor amplitude relative to the
tip of a tethered sub-unit. The form factor amplitude represents the
pair distribution between the reference point at the tip of a tethered
sub-unit and the sites in the same sub-unit, the sites in the core,
and in the sites in the other tethered sub-units. These are shown
in fig. \ref{fig:micelleAP} and when taking into account the multiplicity
of the sub-units the normalized form factor becomes

\[
A_{mic,\omega}(q)=\left(N\beta_{T}+\beta_{C}\right)^{-1}\left(\beta_{C}\Psi_{T,\omega\alpha}A_{C,\langle\alpha\rangle}\right.
\]

\begin{equation}
\left.+\beta_{T}A_{T\omega}+\beta_{T}(N-1)\Psi_{T\omega\alpha}\psi_{C\langle\alpha\rangle\langle\alpha\rangle}A_{T\alpha}\right).\label{eq:a_micelle}
\end{equation}

The contribution to the tip-to-tip phase factor is also shown in fig.
\ref{fig:micelleAP} and is given by

\begin{equation}
\Psi_{mic,\omega\omega}(q)=\psi_{T\alpha\omega}^{2}\Psi_{C\langle\alpha\rangle\langle\alpha\rangle}.\label{eq:p_micelle}
\end{equation}

Since exactly the same diagrams are required to describe a bottle-brush
structure where side-structures are randomly tethered along some main
chain structure (corresponding to the core of the micelle), the generic
scattering expressions for these two structures are identical. They
will first differ when we make choices of which sub-unit structures
are involved and insert the corresponding form factor, form factor
amplitudes, and phase factors in the expressions. 

\begin{figure}
\includegraphics[scale=0.4]{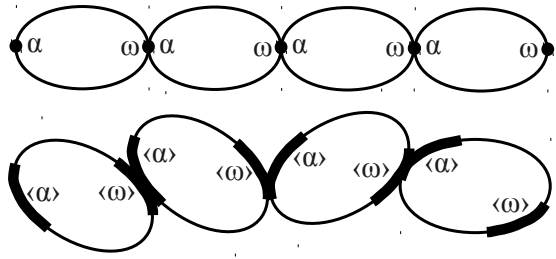}\caption{Diagrammatic representation of chain of identical two-functional sub-units
with regular reference points (top) or linked at two random positions
(bottom).\label{fig:Chain}}
\end{figure}

Previously\cite{cs_jpc_submitted1}, we derived the form factor of
a chain of identical two functional sub-units using eq. \ref{eq:F}
as

\selectlanguage{american}%
\begin{equation}
F_{chain}(q)=\frac{F}{N}+2\frac{\Psi_{\alpha\omega}^{N}-N\Psi_{\alpha\omega}+N-1}{N^{2}(\Psi_{\text{\ensuremath{\alpha\omega}}}-1)^{2}}A_{\alpha}A_{\omega}.\label{eq:f_chain}
\end{equation}

\selectlanguage{english}%
Here we have discarded the superfluous sub-unit index, and also omitted
the $q$ dependence on the right hand side for sake of brevity. A
diagrammatic representation is shown in fig. \ref{fig:Chain}. $\alpha$
and $\omega$ denotes the two ends of the sub-unit. The sub-units
can be asymmetric with regard to exchanging the two ends, for instance
if the sub-unit is a di-block copolymer. The sub-units are joined
as $\omega-\alpha$, leaving one free $\alpha$ end and one free $\omega$
end of the structure. If we assume that each of the reference points
are picked from two distributions, then we obtain the scattering expression
for the corresponding randomly joined chain by replacing the regular
reference points by distributed reference points as 

\begin{equation}
F_{chain}(q)=\frac{F}{N}+2\frac{\Psi_{\langle\alpha\rangle\langle\omega\rangle}^{N}-N\Psi_{\langle\alpha\rangle\langle\omega\rangle}+N-1}{N^{2}(\Psi_{\langle\alpha\rangle\langle\omega\rangle}-1)^{2}}A_{\langle\alpha\rangle}A_{\langle\omega\rangle}.\label{eq:f_chain_random}
\end{equation}

Fig. \ref{fig:Chain} shows a diagrammatical representation of such
a structure, where a random reference point $\omega$ is joined with
a random reference point $\alpha$ on the next sub-unit. Again this
leaves a structure with two ends characterized by $\langle\alpha\rangle$
and $\langle\omega\rangle$. If the sub-units are block-copolymers
and the linking can be anywhere along the copolymer, then the corresponding
structure is one where each polymer is randomly cross-linked with
the previous and next polymers in the chain. Alternative, if the $\alpha$
link is anywhere in the $A$ block, and the $\omega$ link anywhere
in the $B$ block, then the result will be a chain where each di-block
copolymer (except for the ends) has one link on the each of the two
blocks. These different choices correspond to different expressions
for $\Psi_{\langle\alpha\rangle\langle\omega\rangle}$, $A_{\langle\alpha\rangle}$,
and $A_{\langle\omega\rangle}$.

Note that the formalism is generic. We have made absolutely no assumptions
as to the internal structure of the sub-units in the expressions above.
These scattering expressions we have presented are completely generic
and only encode the structural connectivity. The formalism is also
complete, in the sense that a whole structure described by the formalism
can be used as a sub-unit to build more complex structures within
the formalism . For example, we can regard the three functions $F_{mic}(q)$,
$A_{mic,\omega}(q)$, and $\Psi_{mic,\omega\omega}(q)$ given by eqs.
\ref{eq:f_micelle}-\ref{eq:p_micelle} as defining a micelle sub-unit.
We could then obtain the scattering expression for a chain of micelles,
by inserting the micelle sub-unit expressions into the form factor
of a chain eq. \ref{eq:f_chain}. This illustrates the versatility
of the formalism.

\section{Sub-units with internal conformations\label{sec:Linear}}

Each sub-unit is characterized by a three types of pair-distance distribution
functions, the site-to-site, site-to-reference, and the reference-to-reference
point pair-distribution functions, denoted $P_{ss}(\sigma,\rho;r)$,
$P_{s\alpha}(\sigma;r)$, and $P_{\alpha\omega}(r)$, respectively.
Here $\sigma$ and $\rho$ are running labels denoting scattering
sites such as e.g. an index of a point scatterer or a contour-length,
surface or volume element, respectively, while the $\alpha$ and $\omega$
labels denotes fixed reference points. The corresponding positions
are denoted ${\bf r}_{\sigma}$, ${\bf r}_{\rho}$, ${\bf R}_{\alpha}$,
and ${\bf R}_{\omega}$, respectively. In a rigid structure, the pair-distance
$r=|{\bf r}_{\sigma}-{\bf r}_{\rho}|$ is constant and the pair-distance
distributions reduce to delta functions. In a flexible structure with
internal conformational degrees of freedom such as a polymer, the
distance between two sites will in general be given by a distribution.
Let the excess scattering length density of a scattering site $\sigma$
be denoted $b(\sigma)$, and $\beta=\int\mbox{d}\sigma b(\sigma)$
denotes the total excess scattering length of the sub-unit. The 3D
isotropically averaged Fourier transform is $\mathcal{F}(P)=\int\mbox{d}r4\pi r^{2}\frac{\sin(qr)}{qr}P$.
Hence the sub-unit scattering expressions are

\begin{equation}
F(q)=\beta^{-2}\int\mbox{d}\sigma\mbox{d}\rho b(\sigma)b(\rho)\int\mbox{d}r4\pi r^{2}\frac{\sin(qr)}{qr}P_{ss}(\sigma,\rho;r),\label{eq:FormFactor}
\end{equation}

\begin{equation}
A_{\alpha}(q)=\beta^{-1}\int\mbox{d}\sigma b(\sigma)\int\mbox{d}r4\pi r^{2}\frac{\sin(qr)}{qr}P_{s\alpha}(\sigma;r),\label{eq:FormFactorAmplitude}
\end{equation}

\begin{equation}
\Psi_{\text{\ensuremath{\alpha\omega}}}(q)=\int\mbox{d}r4\pi r^{2}\frac{\sin(qr)}{qr}P_{\alpha\omega}(r).\label{eq:PhaseFactor}
\end{equation}

In the case of a linear sub-unit with translational invariance along
the contour, the pair-distance distributions functions only depend
on the relative contour distance. Let $L$ denote the total contour
length, such that $\sigma,\rho\in[0;L]$ denote a pair of sites along
the sub-unit separated by a contour length distance $l=\text{|\ensuremath{\sigma}-\ensuremath{\rho}|}$
and a spatial separation $r$. The two ends are located at $\alpha=0$
and $\omega=L$, respectively. Then $P_{ss}(\sigma,\rho;r)=P(|\sigma-\rho|;r)$,
$P_{s\alpha}(\sigma;r)=P(|\sigma-\alpha|;r)$, and $P_{\alpha\omega}(r)=P(|\alpha-\omega|;r)$
where $P(l;r)$ denotes the pair-distance distribution between two
sites separated by a contour length $l$ and a direct distance $r$.
In Appendix \ref{sec:Appendix-liniear}, we use these expressions
to derive the form factor, form factor amplitudes, and phase factors
of polymers, rods, and closed polymeric loops.

\section{Scattering examples\label{sec:Linear-examples}}

\begin{figure}
\includegraphics[scale=0.5]{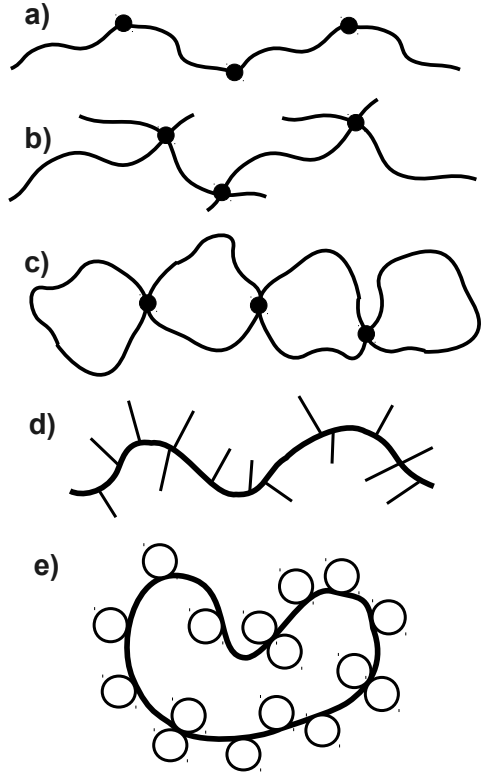}\caption{\label{fig:liniearillustration}Illustration of chains and tethered
structures. a) end-linked polymers, b) contour-linked polymers, c)
contour-linked loops, d) rods contour-linked to a polymer, and e)
loops contour-linked to a loop.}
\end{figure}

\begin{figure}
\includegraphics[scale=0.5]{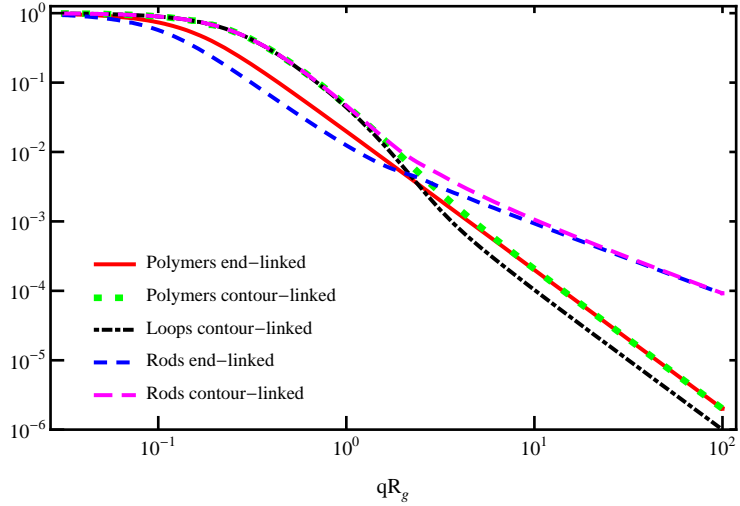}\caption{\label{fig:chain}Normalized form factors from a chain of $N=100$
identical sub-units for end-linked polymers (red solid line), contour-linked
polymers (green dashed), contour-linked polymer loops (blue dotted),
end-linked rods (magenta short dashed), contour-linked rods (brown
medium dashed). All sub-units has the same radius of gyration $R_{g}$.}
\end{figure}

\begin{figure}
\includegraphics[scale=0.5]{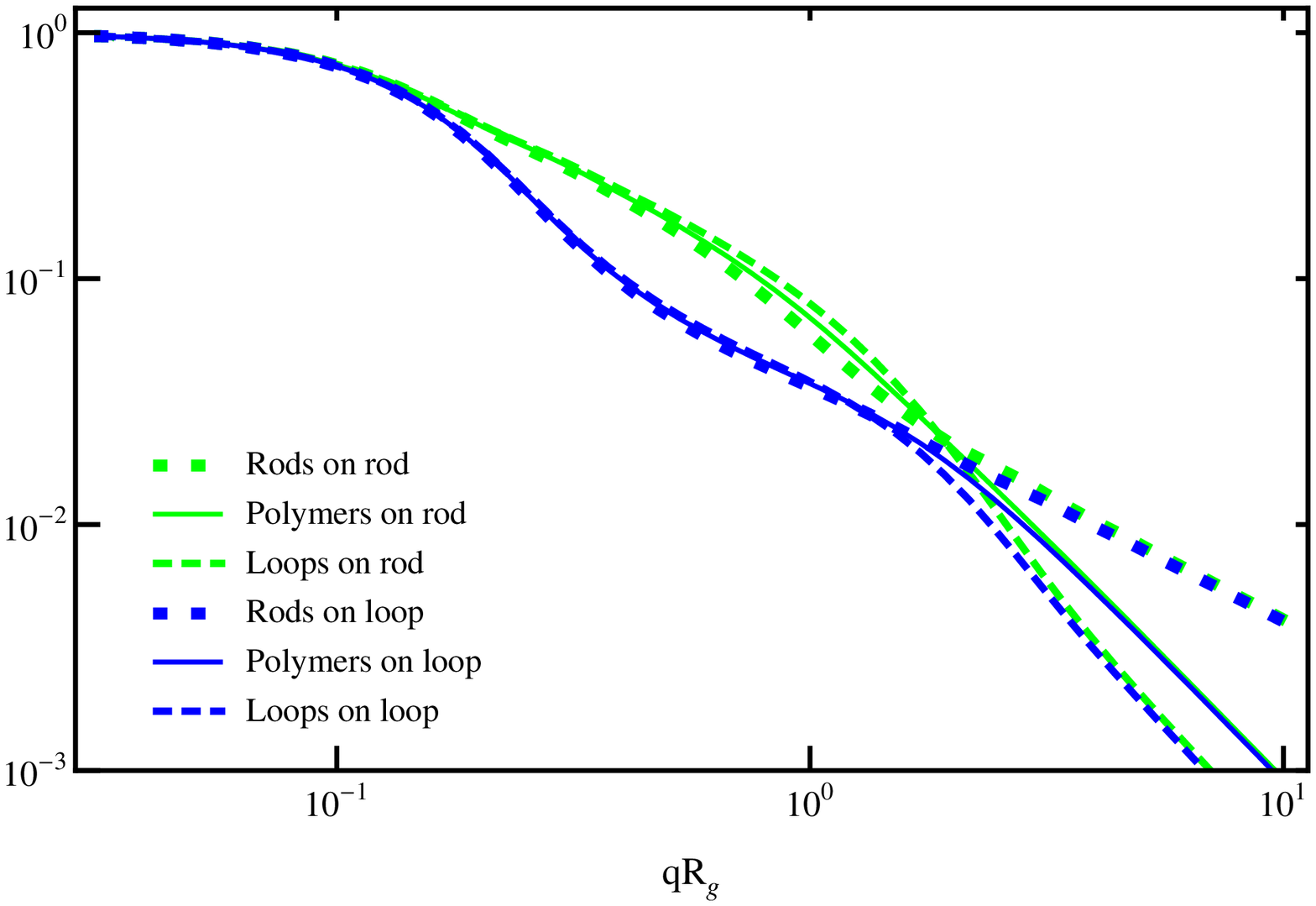}\caption{\label{fig:bottlebrush}Normalised form factor for bottle brush structures
where the main chain structure is either a rod (green) or a polymer
loop (blue). The main chain structure has $N=20$ rods (dotted), polymers
(solid), or polymer loops (dashed) tethered at random positions. The
main chain and tethered sub-unit radii of gyration, $R_{g,C}$ and
$R_{g}$, are fixed at $R_{g,C}/R_{g}=10$ for all the structures.
The excess scattering length of the core and tethered sub-units are
the same $\beta_{C}=\beta_{T}$.}
\end{figure}
Fig. \ref{fig:liniearillustration} a-c illustrates some of the possible
structures obtained by linking sub-units into chains. When identical
polymers are end-linked the result is a long linear polymer. A very
different structure is obtained, when polymers are allows to link
anywhere along their contour. The result resembles a bottle-brush
structure, where each sub-unit in the chain has two pendant chains
of a random length. Note that just as in the end-linked case, all
the internal sub-units in the contour-linked chain have exactly two
links. This is very different from a truly randomly linked structure,
which would form a gel-like network. The scattering from a gel-like
network can be described by the Random Phase Approximation (RPA)\cite{Benoit},
and the diagrammatic representation of the RPA form factor corresponds
to a weighed sum over contour-linked chain diagrams of varying number
of sub-units.

The scattering from these structures are obtained from eq. \ref{eq:f_chain_random}
by inserting the corresponding sub-unit form factor, form factor amplitude
and phase factors. We have derived these terms for a flexible polymer
chain, a rod, and a closed polymer loop in appendix \ref{sec:Appendix-liniear}.
Fig. \ref{fig:chain} shows the scattering from end-linked and contour-linked
polymers and rods, as well as contour-linked loops. At small $q$
values we observe the Guinier $F(q)\approx1+\frac{(qR_{g})^{2}}{3}$
regime, where $R_{g}$ is the radius of gyration of the entire structure,
at intermediate $q$ values we see the power law characteristic of
the fractal dimension of the structure, while at large $q$ values
we see the scattering from the sub-units. The end-linked chain shows
the expected Debye scattering behavior corresponding to a random walk
with an asymptotic behavior $2(qR_{g}){}^{-2}$ at large $q$ values.
The contour-linked polymers have a smaller radius of gyration since
the chain structure is more at intermediate length scales, however
at small length scales we again see the same sub-unit scattering as
for the end-linked polymers. The chain of polymer loops is more compact
than the chain of end-linked polymers, which is why their scattering
is larger at large $q$ values, however at large $q$ values we observe
the asymptotic $(qR_{g})^{-2}$ behaviour expected for a polymer loop.
The end-linked chain of rods is observed to have the largest radius
of gyration of all structures. At intermediate length scales the end-linked
rod chain has a random-walk like structure, while at small length
scales shows a cross-over to the $\pi(Lq)^{-1}$ asymptotic behavior
characterizing the rod-like sub-units. Chains of contour-linked polymers,
polymer loops, and rods have the same radius of gyration, since we
have fixed the size of the sub-units to produce the same radius of
gyration.

Fig. \ref{fig:liniearillustration}d-e shows the bottle-brush structures
that are obtained by tethering rods to a long main chain polymer and
by tethering polymer loops to a main chain polymer loop. The scattering
from these structures is obtained from eq. \ref{eq:f_micelle} by
inserting the corresponding form factor, form factor amplitudes and
phase factors of the main chain and tethered sub-units. The scattering
form factors are shown in fig. \ref{fig:bottlebrush}. Again we have
fixed the radius of gyration of the main chain and of the tethered
sub-units to the same values independent of their structure and for
this reason all the bottle-brush structures has the same radius of
gyration. At intermediate length scales we see a small regime with
power law behavior corresponding to the fractal dimension of the main
chain $q^{-2}$ for the random-walk like polymer loop and $q^{-1}$
for the straight rod, while at small length scales we observe the
power law corresponding to the fractal dimensions of the tethered
sub-units. Again we observe that the polymer loop sub-unit scattering
is a factor one half lower than that of the linear polymer sub-unit.

\section{Geometric sub-units\label{sec:Geometric}}

We assume that a sub-unit is a rigid geometric body without internal
degrees of freedom. In this case, it is more convenient to express
the sub-unit scattering expressions as the orientational average of
the phase integral over all the scattering sites as

\begin{equation}
F_{rigid}(q)=\left\langle {\cal F}_{\beta}({\bf q},{\bf O}){\cal F}_{\beta}(-{\bf q},{\bf O})\right\rangle _{o},\quad A_{rigid}(q,{\bf O})=\left\langle {\cal F}_{\beta}({\bf q},{\bf O})\right\rangle _{o},\label{eq:FA_solid}
\end{equation}
here $\left\langle \cdots\right\rangle _{o}$ denotes an orientational
average. While the form factor is independent of the choice of origin
${\bf O}$, it is useful when expressing form factor amplitudes, since
we have $A_{rigid,\alpha}(q)=A_{rigid}(q,{\bf R}_{\alpha})$ for a
particular regular reference point ${\bf R}_{\alpha}$. The phase
integral is defined as

\begin{equation}
{\cal F}_{\beta}({\bf q},{\bf O})=\left(\int\mbox{d}{\bf r}\beta({\bf r})\right)^{-1}\int\mbox{d}{\bf r}\beta({\bf r})\exp\left(i\mathbf{q}\cdot(\mathbf{r}-\mathbf{O})\right),\label{eq:PhaseIntegral}
\end{equation}
which is the Fourier transform of the excess scattering length density
distribution $\beta({\bf r})$ of the sub-unit relative to the origin
${\bf O}$. We normalize the phase integral such that ${\cal F}_{\beta}({\bf q}=0,{\bf O})=1$.
The major challenge when calculating the scattering from geometric
objects is to calculate the phase integral analytically and then perform
the orientational averages.

Since we are not only interested in regular reference points, but
also reference points that are averaged over distributions of potential
reference point sites, we will focus on the situation where these
site distributions are also characterized by a geometric objects.
For instance, we could be interested in the form factor amplitude
of a sphere relative to a random site on its surface, or the phase
factor between two random sites on the surface of a sphere. By generalizing
the averages eqs. (\ref{eq:aaverage} and \ref{eq:paverage}) into
integrals over reference point distributions, and recognizing that
these averages can be recast into the form of phase factor integrals,
we can express the reference point distribution averaged form factor
amplitude and phase factors as

\begin{equation}
A_{rigid\langle\alpha\rangle}(q)=\left\langle \int\mbox{d}{\bf r}'Q_{\alpha}({\bf r}'){\cal F}_{\beta}({\bf q};{\bf r}')\right\rangle _{o}=\left\langle {\cal F}_{\beta}({\bf q},{\bf O}){\cal F}_{Q_{\alpha}}(-{\bf q},{\bf O})\right\rangle _{o},\label{eq:A_avg}
\end{equation}

\begin{equation}
\Psi_{rigid\langle\alpha\rangle\omega}(q)=\left\langle \int\mbox{d}{\bf r}Q_{\alpha}({\bf r})e^{i{\bf q}\cdot({\bf r}-{\bf R}_{\omega})}\right\rangle _{o}=\left\langle {\cal F}_{Q_{\alpha}}({\bf q},{\bf R}_{\omega})\right\rangle _{o},\label{eq:Pavg}
\end{equation}
and

\begin{equation}
\Psi_{rigid\langle\alpha\rangle\langle\omega\rangle}(q)=\left\langle \int\mbox{d}{\bf r}\mbox{d}{\bf r}'Q_{\alpha}({\bf r})Q_{\omega}({\bf r}')e^{i{\bf q}\cdot({\bf r}-{\bf r}')}\right\rangle _{o}=\left\langle {\cal F}_{Q_{\alpha}}({\bf q},{\bf O}){\cal F}_{Q_{\omega}}(-{\bf q},{\bf O})\right\rangle _{o}.\label{eq:Pavgavg}
\end{equation}

Here ${\cal F}_{Q_{\alpha}}({\bf q},{\bf O})$ denotes the Fourier
transform (eq. \ref{eq:PhaseIntegral}) of the reference point site
probability distribution $Q_{\alpha}({\bf r})$. The form factor amplitude
$A_{\langle\alpha\rangle}(q)$ and double averaged phase factor $\Psi_{\langle\alpha\rangle\langle\omega\rangle}(q)$
are independent of the choice of origin ${\bf O}$ by construction.
Again we recognize that if the normalized excess scattering length
distribution and the reference point site probability distribution
are proportional $\beta({\bf r})\propto Q_{\alpha}({\bf r})$, then
the form factor, averaged form factor amplitude, and double averaged
phase factor reduce to the same function. Finally, the phase factor
between two regular reference points ${\bf R}_{\alpha}$ and ${\bf R}_{\omega}$
is given by 

\begin{equation}
\Psi_{rigid,\alpha\omega}(q)=\frac{\sin(q|\mathbf{R}_{\alpha}-\mathbf{R}_{\omega}|)}{q|\mathbf{R}_{\alpha}-\mathbf{R}_{\omega}|}.\label{eq:P_solid}
\end{equation}

For a large number of geometric objects the scattering form factor
is known, see e.g. \cite{JanAnalysis3}. In an appendix, we derive
the scattering expressions characterizing spheres, flat disks, spheres,
and cylinders with special emphasis on how phase factors and form
factor amplitudes change depending on different choices of reference
point distributions. This is relevant in many applications e.g. for
structures such as block-copolymer micelles and polymer brushes end-grafted
to a surface or an interface.\cite{PedersenGerstenberg,PedersenGerstenberg2}
Below we give some examples.

\section{Scattering examples\label{sec:geometric-examples}}

\begin{figure}
\includegraphics[scale=0.5]{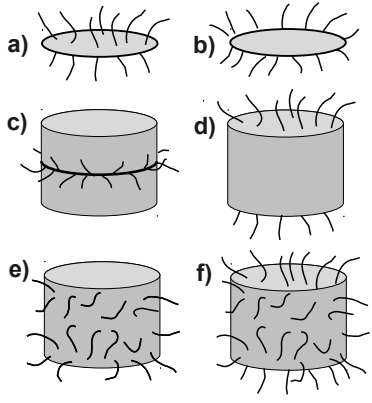}\caption{\label{fig:geometricillustration}Illustration of tethering geometries:
a) disk surface, b) disk rim, c) cylinder equator, d) end, e) side,
and f) surface tethering.}
\end{figure}

\begin{figure}
\includegraphics[scale=0.5]{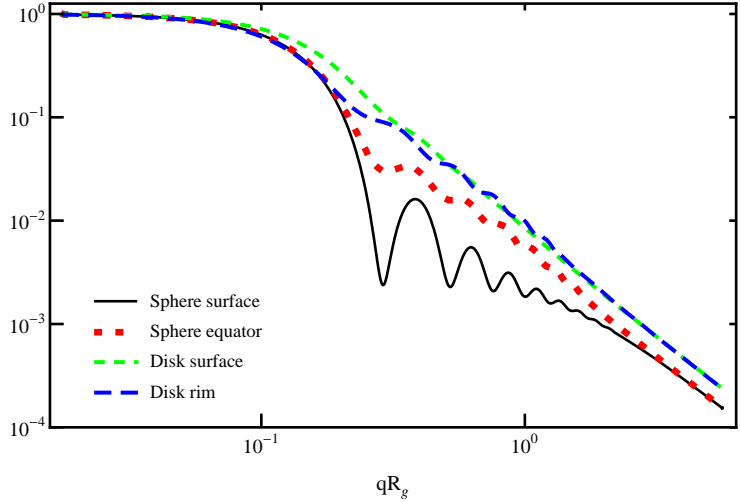}\caption{Normalized form factors for spherical core sub-unit with $N=100$
polymers tethered on the surface (black solid line), at the equator
(red dotted line), and for a disk-like core sub-unit with polymers
tethered on the surface (green short dashed line) and on the rim (blue
long dashed line). The sub-units has the same radii of gyration with
$R_{g,C}/R_{g}=10$, and the excess scattering lengths $\beta_{C}=100\beta_{T}$
.\label{fig:disk_sphere}}
\end{figure}
\begin{figure}
\includegraphics[scale=0.5]{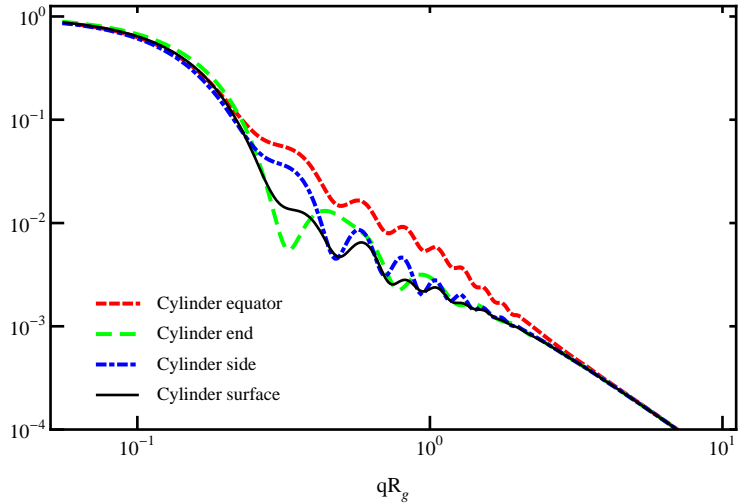}\caption{Normalized form factors of cylinders with $N=100$ polymer sub-units
for equator (black solid line), end (red dotted line), side (green
short dashed line), and surface (blue long dashed line) tethering
geometries. The sub-units has fixed radii of gyration with $R_{g,C}/R_{g}=10$,
the length $L$ and radius $R$ of the cylinder are equal, and the
excess scattering lengths are $\beta_{C}=100\beta_{T}$.\label{fig:cylinder}}
\end{figure}
\begin{figure}
\includegraphics[scale=0.5]{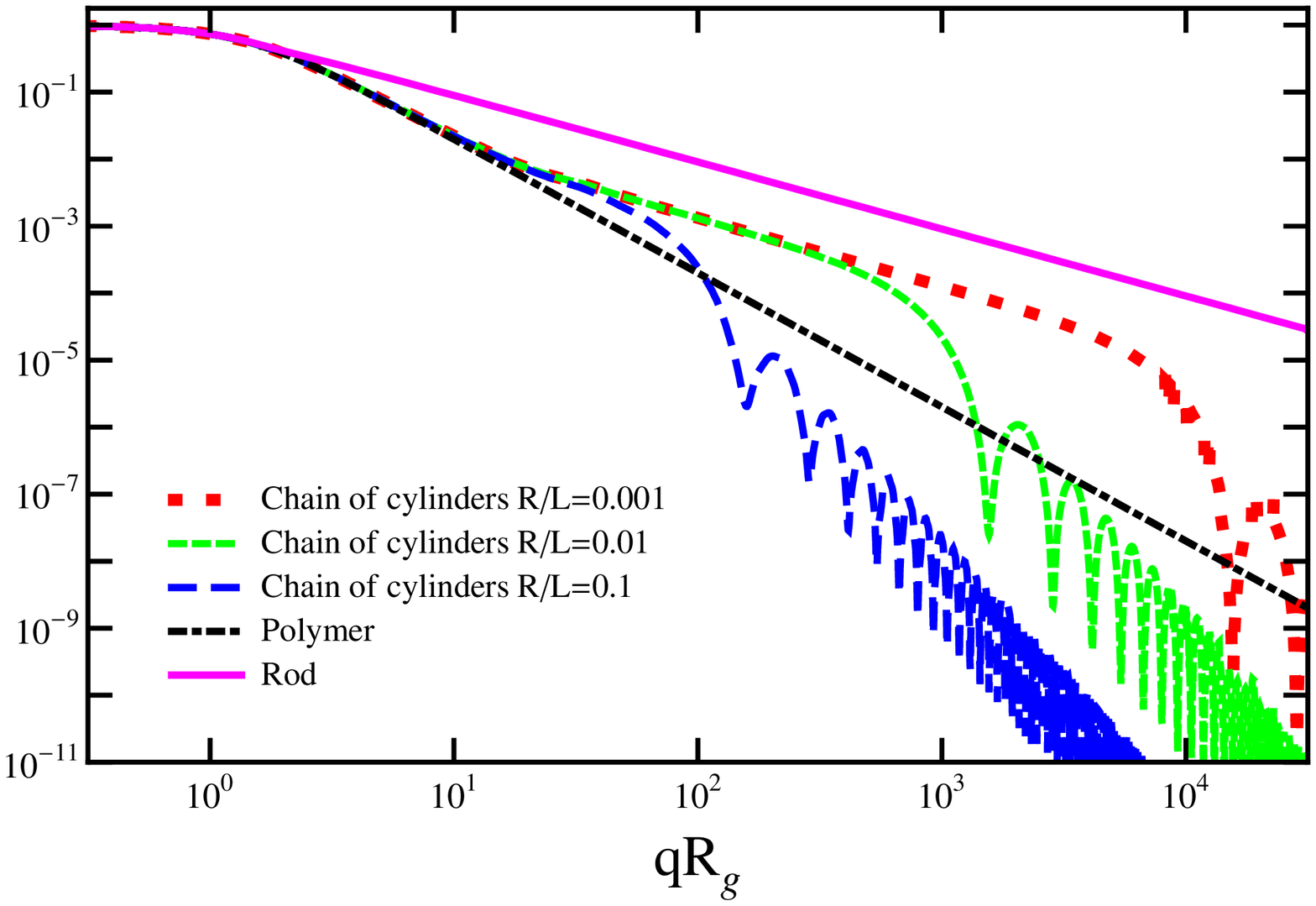}\caption{Normalised form factors for chains of $N=100$ cylinders as function
of their aspect ratio for $R/L=0.001$ (red dotted line), $0.01$
(green short dashed), $0.1$ (blue long dashed). Also shown are the
scattering from a polymer (black dot-dashed) and a rod (magenta solid
line). The horizontal axis of each form factor has been scaled with
the radius of gyration of each of the structures.\label{fig:thickcylinderwalk}}
\end{figure}

Fig. \ref{fig:geometricillustration} shows some of the possible tethering
geometries for disk-like and cylindrical micelles. For a disk we can
either have the sub-units tethered to anywhere on the surface, or
only at the rim of the surface. For a cylinder we could for instance
tether polymers to the equator, the two ends, the side, or the entire
surface of the cylinder. The corresponding scattering expressions
are obtained from eq. \ref{eq:f_micelle} by inserting the form factor,
form factor amplitudes, and phase factors corresponding to the chosen
sub-units and tethering geometry. In Appendix \ref{sec:Appendix-geometric},
we have derived the required expressions to characterize these tethering
geometries.

Fig. \ref{fig:disk_sphere} shows the form factors for disk-like and
spherical micelles. At small $q$ values we observe the Guinier regime
which characterizes the radius of gyration of the whole structure,
while at large $q$ values we observe the scattering due to the tethered
chain sub-unit. In an intermediate regime, the scattering is both
due to the micellar core geometry and the tethered sub-units. Even
though the number of chains is the same, significant differences are
observed in the scattering for the different tethering geometries,
but coincidentally the sphere with equatorial tethering and disk with
rim tethering produce very similar scattering patterns. Fig. \ref{fig:cylinder}
shows the form factors for cylindrical micelles for different choices
of tethering geometry. Again the tethering geometry is observed to
introduce significant differences in the spectra.

The scattering expression for a chain of thick end-linked cylinders
are obtained from eq. \ref{eq:f_chain} by inserting the form factor,
form factor amplitude relative to the reference point where the axis
crosses the end, and phase factor between the two ends. In fig. \ref{fig:thickcylinderwalk},
the scattering from this thick random walk is compared to that of
a thin polymer and a rod. At large distances in the Guinier regime
we see the crossover from a point like structure at the very largest
scales to a random walk like structure with scaling behavior $(qR_{g})^{-2}$.
At intermediate length scales the cylinder structure is probed and
shows a scaling behavior like $(qL)^{-1}$ comparable to the rod.
At length scales at and below the radius of the rod, we see strong
oscillations due to cross section of the cylinder, and the envelope
of the scattering curve shows the $q^{-4}S^{2}$ behavior of Porod
scattering from the surface, where $S$ denotes the total surface
area of the cylinders.

\section{Conclusions\label{sec:Conclusions}}

In a previous paper\cite{cs_jpc_submitted1}, we presented a formalism
for predicting the scattering from a linear and branched structures
composed of mutually non-interacting sub-units. A sub-unit can have
an arbitrary number of reference points. Sub-units are connected into
structure by joining their reference points to the reference points
of other sub-units. In the present paper, we have briefly presented
the formalism, and generalized it to the case where reference points
can be characterized by a distribution of potential link positions
on a sub-unit. For instance, one reference point of a polymer could
be a random site along the contour, or a reference point of a sphere
could be a random point on the surface. To generalize the formalism,
we had to assume that reference point distributions on different sub-units
are mutually statistically uncorrelated.

We used the generalized formalism to derive the generic scattering
expression for a micelle / bottle-brush structure with a core sub-unit
and a number of identical sub-units tethered to random positions on
the core / main chain. Since the connectivity of a micelle and a bottle-brush
is the same, the generic scattering expressions are also identical.
We presented the general scattering expressions for the form factor,
form factor amplitudes and phase factors of a structure with internal
conformations. We illustrated the scattering the expression using
end-linked and contour-linked chains of polymers, rods, and polymeric
loops and bottle-brush structures of rods and polymer loops with tethered
polymers, rods, or polymer loops. All these structures are special
cases of the generic scattering expression, which are obtained when
the form factor, form factor amplitudes, and phase factors of the
corresponding sub-units are inserted into the generic structural scattering
expression. We derived these terms in an appendix.

We also presented the general expressions for the form factor, form
factor amplitudes, and phase factors for rigid sub-units without internal
conformations. We derived the scattering terms for spheres, disks,
and cyllinders for a variety of different reference point distributions
in an appendix. While the form factors for all these sub-units are
known, the form factor amplitudes and phase factors are not necessarily
known, and these are required by the formalism. This allowed us to
predict the scattering from micelles with different core structures
and geometries of tethering the corona sub-units.

Taken together, the formalism presented in ref. \cite{cs_jpc_submitted1},
the present generalization to distributed reference points, and the
sub-unit scattering expressions derived in the appendix enables the
scattering from a large class of regular or random-linked, homogeneous
or heterogeneous, linear and branched structures to be derived with
great ease. With this, we hope to have provided a valuable tool for
analyzing scattering data in the future.

\section{Acknowledgments}

C.S. and J.S.P. gratefully acknowledges discussions with C.L.P. Oliveira.
Funding for this work is provided in part by the Danish National Research
Foundation through the Center for Fundamental Living Technology (FLinT).
The research leading to these results has received funding from the
European Community's Seventh Framework Programme (FP7/2007-2013) under
grant agreement n\textdegree{} 249032, MatchIT - Matrix for Chemical
IT.

\section{Appendix: Sub-units with internal conformations\label{sec:Appendix-liniear}}

In the following we derive the scattering expressions for rigid rods,
flexible polymers, and closed polymer loops using eqs. \ref{eq:FormFactor}-\ref{eq:PhaseFactor}.

\subsection{Rigid rods\label{sec:Rigid-rods}}

The most basic example is a randomly orientated infinitely thin rigid
rod of length $L$. The rigid rod is special as the contour length
$l$ and direct distance $r$ between a pair sites are degenerate
parameters, hence the $\delta(r-l)$ factor in the rod pair-distance
distribution: $P(l;r)=\delta(r-l)\Theta(L-r)/(4\pi l^{2})$, where
$\Theta(x)$ denotes the Heaviside step function. Performing the contour
length integrations (\ref{eq:FormFactor})-(\ref{eq:PhaseFactor}),
the rod scattering triplet becomes
\begin{equation}
F_{rod}(q,L)=\frac{2\mbox{Si}(x)}{x}-\frac{4}{x^{2}}\sin^{2}\left(\frac{x}{2}\right),\label{eq:rod1}
\end{equation}
\begin{equation}
A_{rod}(q,L)=\frac{\mbox{Si}(x)}{x}\quad\Psi_{rod}(q,L)=\frac{\sin(x)}{x},\label{eq:rod2}
\end{equation}
where $x=qL$ and $\mbox{Si}(x)=\int_{0}^{x}\mbox{d}y\sin(y)/y$ is
the Sin integral\cite{AbramowitzStegun}. The expression for the rod
form factor was previous derived by Neugebauer\cite{Neugebauer} and
Teixera et al \cite{TeixeiraJCP2007}. In the case where the reference
points are distributed along the rod contour with a constant probability
(denoted contour-linking and shown with sub-script $\langle c\rangle$
and $\langle c\rangle\langle c\rangle$), the rod phase factor and
form factor amplitudes are given by 

\[
\Psi_{rod,\langle c\rangle\langle c\rangle}(q)=A_{rod,\langle c\rangle}(q)=F_{rod}(q,L).
\]

\subsection{Flexible polymers\label{sub:Flexible-polymers}}

Flexible polymers can be modeled as random walks with a effective
step length or Kuhn length $b$. The pair-distance distributions between
two sites separated by a contour length $l$ are given by the Gaussian
distribution 
\[
P_{rw}(l;r)=\left(\frac{3}{2\pi bl}\right)^{\frac{3}{2}}\exp\left(-\frac{3}{2}\frac{r^{2}}{bl}\right)
\]

Inserting this distribution into eqs. (\ref{eq:FormFactor})-(\ref{eq:PhaseFactor})
yields the scattering triplet characterizing a flexible polymer 
\begin{equation}
F_{pol}(x)=\frac{2[e^{-x}-1+x]}{x^{2}},\quad A_{pol}(x)=\frac{1-e^{-x}}{x},\quad\mbox{and}\quad\Psi_{pol}(x)=e^{-x}\label{eq:flex}
\end{equation}
here $x=(qR_{g,rw})^{2}$, and the radius of gyration is given by
$R_{g,rw}^{2}=bL/6$. The result for the form factor amplitude has
previously been given by Hammouda \cite{Hammouda} and the form factor
was calculated by Debye \cite{Debye}. These expressions can also
be obtained as a self-consistency requirement of this formalism requiring
that the scattering form factor, form factor amplitude and phase factor
from a di-block copolymer with two identical blocks of length $L/2$
match the scattering expressions for one block of length $L$.\cite{cs_jpc_submitted1}
Expressions for the scattering from poly disperse flexible polymer
characterized by a Schultz-Zimm distribution are given in Ref. \cite{BenoitHadziioannouMacromolecules88}.
In the case where the reference points are distributed randomly along
the polymer contour (denoted $\langle c\rangle$ and $\langle c\rangle\langle c\rangle$),
the polymer phase factor and form factor amplitude are given by 

\[
\Psi_{pol,\langle c\rangle\langle c\rangle}(x)=A_{pol,\langle c\rangle}(x)=F_{pol}(x).
\]

\subsection{Polymer loops\label{sub:Polymer-loops}}

While the formalism does not apply for structures that contains loops,
no assumptions are made as to the internal structure of the sub-units,
which can contain loops. The simplest loop is formed by linking the
two ends of a flexible polymer chain. We can model a polymer loop
as two random walks with contour lengths $l$ and $L-l$, respectively,
starting at ${\bf R}_{\alpha}$ and ending at the link $\mathbf{R}_{\omega}$.
In this case, the pair-distance distribution is given by

\[
P_{loop}(l;|\mathbf{R}_{\omega}-\mathbf{R}_{\alpha}|)\propto P_{rw}(l;|\mathbf{R}_{\omega}-\mathbf{R}_{\alpha}|)P_{rw}(L-l;|\mathbf{R}_{\omega}-\mathbf{R}_{\alpha}|),
\]
and the corresponding phase factor becomes

\[
\Psi_{loop,\alpha\omega}(q;l)=\exp\left(-\frac{bl(L-l)q^{2}}{6L}\right).
\]

Since the link at $l$ can be anywhere along the loop $[0;L]$, we
need to average over the link position to get the loop scattering: 

\[
\Psi_{loop,\langle\alpha\rangle\langle\omega\rangle}(q)=\int_{0}^{L}\frac{\mbox{d}l}{L}\Psi_{loop}(q;l)=\frac{\exp\left(-2y^{2}\right)D[y]}{y}.
\]

Here $y=q\sqrt{bL}/\sqrt{24}$ and $D[y]=\exp(-y^{2})\int_{0}^{y}\exp(t^{2})dt$
is the Dawson integral, which is related to the imaginary part of
the complex error function\cite{AbramowitzStegun}. By construction,
the form factor amplitude, form factor and average phase factor are
all identical when the reference point(s) is averaged over all sites
in the structure, since eqs. (\ref{eq:FormFactor})-(\ref{eq:PhaseFactor})
become identical $A_{loop,\langle\alpha\rangle}(q)=F_{loop}(q)=\Psi_{loop,\langle\alpha\rangle\langle\omega\rangle}(q)$.
The form factor of a flexible polymer loop was previous derived by
Zimm and Stockmayer.\cite{ZimmStockmayer}

\section{Appendix: geometric sub-units\label{sec:Appendix-geometric}}

Below we will derive the scattering from a solid sphere, a flat disk,
and a cylinder for a variety of reference point distributions using
eqs. \ref{eq:FA_solid}-\ref{eq:Pavgavg}.

\subsection{Solid sphere}

For a solid homogeneous sphere with excess scattering length density
$\beta$, we can characterize any scattering site by its spherical
coordinate $\sigma=(r,\phi,\theta)$. Then $\mathbf{r}(\sigma)=(r\cos\phi\sin\theta,r\sin\phi\sin,r\cos\theta)$.
We can choose a coordinate system such that the sphere center is located
at the origin ${\bf O}=0$. Due to the spherical symmetry the scattering
vector $\mathbf{q}$ is pointing towards the pole ($\theta=0)$, then
$\mathbf{q}\cdot\mathbf{r}_{\sigma}=qr\cos\theta$. The phase integral
becomes

\[
{\cal F}_{sphere}({\bf q},0;R)=\left(\frac{4\pi R^{3}}{3}\right)^{-1}\int_{-\pi}^{\pi}\mbox{d}\phi\int_{-1}^{1}\mbox{d}\cos\theta\int_{0}^{R}\mbox{d}rr^{2}e^{iqr\cos\theta}
\]

\begin{equation}
=\frac{3\left(\sin(qR)-qR\cos(qR)\right)}{(qR)^{3}},\label{eq:sphericalvolume}
\end{equation}

Due to the spherical symmetry, we do not need to perform an additional
orientational average $\langle\cdots\rangle_{o}$ when using eqs.
\ref{eq:FA_solid}-\ref{eq:P_solid}. Hence, the form factor, the
form factor amplitude and phase factor of a solid sphere with $\mathbf{R}_{\alpha}=\mathbf{R}_{\omega}=0$
fixed at the center (denoted subscript {}``c'') are given by

\begin{equation}
F_{sphere}(q;R)=A_{sphere,c}^{2}(q;R),\label{eq:F_solidspherecc}
\end{equation}

\[
A_{sphere,c}(q;R)=\frac{3\left(\sin(qR)-qR\cos(qR)\right)}{(qR)^{3}},
\]

\begin{equation}
\Psi_{sphere,cc}(q)=1.\label{eq:P_solidspherecc}
\end{equation}

The scattering from a solid sphere was derived by Reyleigh in 1911\cite{Rayleigh}.
Having the reference points at the center is the simplest choice,
however, for the derivation of e.g. the scattering from spherical
micelles it is more relevant let the surface of the sphere be a reference
point. In this case the corresponding normalized reference point distributions
are $Q_{\alpha}({\bf R})=Q_{\omega}({\bf R})=\delta(|{\bf R}|-R)/(4\pi R^{2})$
representing a spherical shell. We can calculate ${\cal F}_{shell}({\bf q},0;R)$
by integration of eq. \ref{eq:PhaseIntegral}. However, since the
shell corresponds to the upper limit of the radial integral in eq.
\ref{eq:sphericalvolume}, we can trivially obtain its Fourier transform
by differentiation of ${\cal F}_{sphere}$ as

\begin{equation}
{\cal F}_{shell}({\bf q},0;R)=\left(4\pi R^{2}\right)^{-1}\frac{\mbox{d}}{\mbox{d}R}\left[\frac{4\pi R^{3}}{3}{\cal F}_{sphere}({\bf q},0,R)\right]=\frac{\sin qR}{qR}.\label{eq:Ishell}
\end{equation}

Here we have introduced an area and volume prefactor to account for
the normalizations of the two Fourier transforms, such that ${\cal F}_{shell}\rightarrow1$
when $q\rightarrow0$. We can obtain the form factor amplitude and
phase factors relative to the surface reference point (denoted by
subscript {}``$\langle s\rangle$'') combining eqs. \ref{eq:sphericalvolume},
\ref{eq:Ishell}, \ref{eq:A_avg}, and \ref{eq:Pavgavg} as

\[
A_{sphere,\langle s\rangle}(q,R)=\frac{3\left(\sin(qR)-qR\cos(qR)\right)}{(qR)^{3}}\times\frac{\sin(qR)}{qR},
\]

and

\begin{equation}
\Psi_{sphere,\langle s\rangle\langle s\rangle}(q,R)=\left(\frac{\sin(qR)}{qR}\right)^{2}.\label{P_solidsphere_ss}
\end{equation}

The spherical phase factor was previously derived by Pedersen and
Gerstenberg.\cite{PedersenGerstenberg}

\subsection{Flat circular disk\label{sub:sphericaldisk}}

Due to rotational symmetry, we can choose a geometry where the disk
is in the $xy$ plane, and $\mathbf{q}$ in the $xz$ plane. Expressing
the scattering site in polar coordinates $\sigma=(r,\phi)$, such
that $\mathbf{r}=(r\cos\phi,r\sin\phi,0)$, and $\mathbf{q}=(q\sin\theta,0,q\cos\theta)$
then $\mathbf{q}\cdot\mathbf{r}=qr\cos\phi\sin\theta$.

\begin{equation}
{\cal F}_{disk}({\bf q},0;R)=\frac{2J_{1}(qR\sin\theta)}{qR\sin\theta}\label{eq:Fourier_disk}
\end{equation}
Expressing the integrals in cylindrical coordinates immediately provides
the form factor, form factor amplitude and phase factor for $\mathbf{R}_{\alpha}=\mathbf{R}_{\omega}=0$
fixed at the center as

\[
F_{disk}(q)=\left\langle \left(\frac{2J_{1}(qR\sin\theta)}{qR\sin\theta}\right)^{2}\right\rangle _{o}=\frac{2}{q^{2}R^{2}}\left[1-\frac{J_{1}(2qR)}{qR}\right],
\]

\begin{equation}
A_{disk,c}(q)=\left\langle \frac{2J_{1}(qR\sin\theta)}{qR\sin\theta}\right\rangle _{o},\quad\Psi_{disk,cc}(q)=1\label{eq:FAP_disk_c}
\end{equation}

Here $J_{n}(x)$ denotes the $n$'th Bessel function of the first
kind\cite{AbramowitzStegun}, and $\langle\cdots\rangle_{o}=\frac{1}{2}\int_{-1}^{1}\mbox{d}\cos(\theta)\cdots$
denotes the remaining orientational average, which needs to be performed
numerically. The form factor of a disk first derived by Kratky and
Porod\cite{KratkyPorod1949a}. We could also choose to put the reference
point anywhere on the circular rim of the disk. In this case, we can
again obtain the Fourier transform of the points on a circular rim
${\cal F}_{rim}$ by differentiation of ${\cal F}_{disk}$ as

\[
{\cal F}_{rim}({\bf q},0,R)=\left(2\pi R\right)^{-1}\frac{\mbox{d}}{\mbox{d}R}\left[\pi R^{2}{\cal F}_{disk}({\bf q},0,R)\right]
\]

\begin{equation}
=J_{0}(qR\sin\theta)\label{eq:Fourier_rim}
\end{equation}

The corresponding form factor amplitude of the disk relative to any
site on the rim and the phase factor between two sites on the rim
(denoted by subscript $\langle r\rangle$)

\begin{equation}
A_{disk,\langle r\rangle}(q)=\left\langle \frac{2J_{1}(qR\sin\theta)}{qR\sin\theta}\times J_{0}(qR\sin\theta)\right\rangle _{o},\label{eq:Adisk_r}
\end{equation}

\begin{equation}
\Psi_{disk,\langle r\rangle\langle r\rangle}(q)=\left\langle J_{0}^{2}(qR\sin\theta)\right\rangle _{o}.\label{Pdisk_rr}
\end{equation}

If we instead chooses to put the two reference points at any point
on the surface of the disk (again denoting an average surface reference
point by $\langle s\rangle$), the result is again given by the disk
form factor as we have seen previously:

\begin{equation}
A_{disk,\langle s\rangle}(q)=\Psi_{disk,\langle s\rangle\langle s\rangle}(q)=F_{disk}(q).\label{eq:APF_disk_avgr}
\end{equation}

\subsection{Solid cylinder\label{sub:Cylindrical-surface}}

We can choose a cylinder with its center at the origin and the axis
along $z$, the natural choice is to describe it with polar coordinates
$\sigma=(r,\phi,z)$ such that $\mathbf{r}(\sigma)=(r\cos\phi,r\sin\phi,z)$
and we define $\mathbf{q}=(q\sin\theta,0,q\cos\theta)$ then $\mathbf{q}\cdot\mathbf{r}(\sigma)=qr\cos\phi\sin\theta+qz\cos\theta$.
The phase integral becomes

\begin{equation}
{\cal F}_{cyl}({\bf q},{\bf O};R,L)=\frac{4J_{1}(qR\sin\theta)\sin(\frac{qL}{2}\cos\theta)}{LRq^{2}\sin\theta\cos\theta}\times e^{-i\mathbf{q}\cdot\mathbf{O}}.\label{eq:Phaseintegral_cyl}
\end{equation}

We have chosen to explicitly write the origin ${\bf O}$, since this
will be required to calculate the form factor amplitude. With regular
reference points at the ends of the cylinder axis $\mathbf{R}_{\alpha}=(0,0,-L/2)$
and $\mathbf{R}_{\omega}=(0,0,+L/2)$ (denoted by subscript {}``$a$''),
the form factor, form factor amplitude and phase factor can be derived
as

\begin{equation}
F_{cyl}(q;R,L)=\left\langle \left(\frac{4J_{1}(qR\sin\theta)\sin(\frac{qL}{2}\cos\theta)}{LRq^{2}\sin\theta\cos\theta}\right)^{2}\right\rangle _{o},\label{eq:F_cyl}
\end{equation}

\begin{equation}
A_{cyl,a}(q;R,L)=\left\langle {\cal F}_{cyl}({\bf q},{\bf R}_{\alpha};R,L)\right\rangle _{o}=\left\langle {\cal F}_{cyl}({\bf q},{\bf R}_{\omega};R,L)\right\rangle _{o}\label{eq:}
\end{equation}

\begin{equation}
=\left\langle \frac{4J_{1}(qR\sin\theta)\sin(\frac{qL}{2}\cos\theta)}{LRq^{2}\sin\theta\cos\theta}\cos\left(\frac{qL}{2}\cos\theta\right)\right\rangle _{o},\label{eq:A_cyl}
\end{equation}

\begin{equation}
\Psi_{cyl,aa}(q;R,L)=\frac{\sin qL}{qL}.\label{eq:P_cyl_aa}
\end{equation}

The form factor of a solid cylinder was previously derived by Fournet\cite{Fournet1949}.
We can also derive the Fourier transform of the end and side of the
cylinder by differentiation as

\[
{\cal F}_{cyl,end}({\bf q},0;R,L)=\left(\pi R^{2}\right)^{-1}\frac{\mbox{d}}{\mbox{d}L}\left[\pi LR^{2}{\cal F}_{cyl}({\bf q},0;R,L)\right]
\]

\begin{equation}
=\frac{2J_{1}(qR\sin\Theta)\cos(\frac{Lq}{2}\cos\Theta)}{qR\sin\Theta}\label{eq:Fourier_cyl_end}
\end{equation}

\[
{\cal F}_{cyl,side}({\bf q},0;R,L)=\left(2\pi RL\right)^{-1}\frac{\mbox{d}}{\mbox{d}R}\left[\pi LR^{2}{\cal F}_{cyl}({\bf q},0;R,L)\right]
\]

\begin{equation}
=\frac{2J_{0}(qR\sin\Theta)\sin(\frac{Lq}{2}\cos\Theta)}{Lq\cos\Theta}\label{eq:Fourier_cyl_hull}
\end{equation}

By combining eqs. \ref{eq:Fourier_cyl_end} and \ref{eq:Fourier_cyl_hull}
weighting the terms by their relative areas and normalizing, we obtain
the Fourier transform of the surface of a cylinder as\cite{PedersenGerstenberg2}

\begin{equation}
{\cal F}_{cyl,sur\! face}({\bf q},0;R,L)=(R+L)^{-1}\left(R{\cal F}_{cyl,end}+L{\cal F}_{cyl,side}\right)\label{eq:Fourier_cyl_surface}
\end{equation}

With these Fourier transforms we can write down the form factor amplitudes
and phase factors of the cylinder relative to a reference point distributed
on the ends (denoted by subscript $\langle e\rangle)$, on the hull
(denoted by $\langle h\rangle$) or anywhere on the surface (denoted
by $\langle s\rangle$) as

\[
A_{cyl,\langle e\rangle}(q;R,L)=\left\langle \frac{8J_{1}^{2}(qR\sin\theta)\sin(\frac{qL}{2}\cos\theta)\cos(\frac{Lq}{2}\cos\theta)}{LR^{2}q^{3}\sin^{2}\theta\cos\theta}\right\rangle _{o},
\]

\[
A_{cyl,\langle h\rangle}(q;R,L)=\left\langle \frac{8J_{0}(qR\sin\theta)J_{1}(qR\sin\theta)\sin^{2}(\frac{qL}{2}\cos\theta)}{RL^{2}q^{3}\sin\theta\cos^{2}\theta}\right\rangle _{o},
\]

\[
A_{cyl,\langle s\rangle}(q;R,L)=\left\langle \frac{4J_{1}(qR\sin\theta)\sin(\frac{qL}{2}\cos\theta)}{LRq^{2}\sin\theta\cos\theta}\right.
\]

\[
\left.\times\frac{2}{q(R+L)}\left(\frac{J_{1}(qR\sin\Theta)\cos(\frac{Lq}{2}\cos\Theta)}{\sin\Theta}+\frac{J_{0}(qR\sin\Theta)\sin(\frac{Lq}{2}\cos\Theta)}{\cos\Theta}\right)\right\rangle _{o},
\]

\[
\Psi_{cyl,\langle e\rangle\langle e\rangle}=\left\langle \left(\frac{2J_{1}(qR\sin\Theta)\cos(\frac{Lq}{2}\cos\Theta)}{qR\sin\Theta}\right)^{2}\right\rangle _{o},
\]

\[
\Psi_{cyl,\langle h\rangle\langle h\rangle}=\left\langle \left(\frac{2J_{0}(qR\sin\Theta)\sin(\frac{Lq}{2}\cos\Theta)}{Lq\cos\Theta}\right)^{2}\right\rangle _{o},
\]

and

\[
\Psi_{cyl,\langle s\rangle\langle s\rangle}=\left\langle \frac{4}{q^{2}(R+L)^{2}}\left(\frac{J_{1}(qR\sin\Theta)\cos(\frac{Lq}{2}\cos\Theta)}{\sin\Theta}\right.\right.
\]

\[
\left.\left.+\frac{J_{0}(qR\sin\Theta)\sin(\frac{Lq}{2}\cos\Theta)}{\cos\Theta}\right)^{2}\right\rangle _{o}.
\]

\pagebreak{}

\bibliographystyle{apsrev2}
\bibliography{thesis}

\end{document}